\documentclass[12pt] {article}
\def\m{\multicolumn}
\newfont{\jnp}{cmss10}
  \def\n{\noindent}
      \def\bea{\begin{eqnarray}}
  \def\eea{\end{eqnarray}}    
  \def\be{\begin{equation}}  \def\ee{\end{equation}}
\begin{document}
{\begin{center}

{\Large\bf Radiative decay of single charmed
baryons}\\

{\large Ajay Majethiya$^\dag$ \footnote{ajay.phy@gmail.com}, Bhavin Patel $^\ddag$ and P C Vinodkumar$^\ddag$}\\
{\small  $^\dag$ $^\ddag$Department of Physics, Sardar Patel University,\\
  Vallabh Vidyanagar, Anand, Gujarat-388 120.}\\
 {\small  $^\dag$Kalol Institute of Technology and Research Center,Gujarat University,\\
  Kalol, Gujarat-382 721.}\\

\end{center}}

\begin{abstract}
The electromagnetic transitions between ($J^{P}=\frac{3} {2}^{+}$)
and ($J^{P}=\frac{1} {2}^{+}$)   baryons are important decay modes
to observe new hadronic states experimentally. For the estimation
of these transitions widths, we employ a non-relativistic quark
potential model description with color coulomb plus linear
confinement potential. Such a description has been employed to
compute the ground state masses and magnetic moments of the single
heavy flavor baryons. The magnetic moments of the baryons are
obtained using the spin-flavor structure of the constituting quark
composition of the baryon. Here, we also define an effective
constituent mass of the quarks (ecqm) by taking into account the
binding effects of the quarks within the baryon. The radiative
transition widths are computed in terms of the magnetic moments of
the baryon and the photon energy. Our results are compared with
other theoretical models.
\end{abstract}
\section{Introduction}
 Recent experimental observations of excited charmed baryons by
 Belle and Babar collaborations have generated \cite{Mizuk2005,Aubert2007} an increasing interest on
heavy-baryon spectroscopy. It is striking that baryons containing
one or two heavy charm or beauty flavour could play an important
role in our understanding of QCD at the hadronic scale
\cite{Garcilazo2007}. Apart from spectroscopy, various decay
processes of the heavy flavour baryons are more important to
observe new hadronic states experimentally. The strong decays are
expected to dominate the branching rates of charmed baryons. In
fact, most of the experimentally discovered channels of ground
state charmed baryons are the one- and two- pion transitions
\cite{Brandenburg1997}. Although the electromagnetic strength is
weaker than that of the strong interaction, radiative channels are
not phase space suppressed as in the case of pion transitions.
Therefore, some radiative decay modes are expected to contribute
significantly to some heavy baryon branching fractions. The recent
observations of CLEO confirm the importance of the radiative
channels even though their contributions to some heavy baryon
widths are relatively small compared to other decay modes. Many
theoretical models have predicted the heavy baryon mass spectrum
\cite{Hwang2008,Roberts2007,Amand2006,Giannini2001,Ebert2005,Yu2006,Bhavin2008,
B.Silvestre-brac1996,Santopinto1998,Rosner609}.  Nevertheless, the
new experimental data of the charmed baryons can be used to test
the success and validity of the different phenomenological models
available in the literature. The study of heavy baryons further
provides excellent laboratory to understand the dynamics
of light quarks in the vicinity of heavy flavour quark as bound states.\\
Although radiative decays are well measured in the case of
$D^*\rightarrow D\gamma$, $D_{s}^+\rightarrow D_{s}^+\gamma$, only
very few cases of $\Xi_{c}^{'0}\rightarrow \Xi_{c}^{0}+\gamma$,
$\Xi_{c}^{'+}\rightarrow \Xi_{c}^{+}+\gamma$ and
$\Omega_{c}^{*0}\rightarrow \Omega_{c}^{0}+\gamma$, have been
reported \cite{PDG2006}. It may be noted that the nonrelativistic
quark model predictions for the magnetic moments of ordinary
baryons are in good agreement with their experimental values; in
particular, it gives a reasonable value for the transition rate
$\Sigma^{0}\rightarrow \Lambda^{0}+\gamma$. It is therefore
reasonable to assume that the estimates of the radiative decays
based on the nonrelativistic quark model is reliable. \\In this
paper, we compute the masses and magnetic moments of the single
charmed baryons using coloumb plus linear as the confinement
inter-quark potential in a non-relativistic framework. The
magnetic moments of the baryons are obtained using the spin-flavor
structure of the constituent quark mass (cqm) parameters employed
in the model as well as with an effective constituent mass of the
quarks (ecqm) by taking into account the binding effects of the
quarks constituting the baryon. We use the present values of
magnetic moments of charmed baryons to obtain the radiative
transition widths.

\section{Methodology}
\begin{table*}
\begin{center}
\caption{Radiative decay widths ($\Gamma_{\gamma}$) of singly
charmed baryons in terms of KeV (* indicates $J^P=\frac{3}{2}^+$
state.)}\vspace{0.001in} \label{tab:01}

\item[]
\begin{tabular}{@{}lrrrrrr}
\hline Decay&$\mu$ (in $\mu_{N}$) &&Present{\,\,\,\,\,\,\,\,\,}&&HCM{\,\,\,\,\,\,\,\,\,\,\,\,\,\,}&Others\\
\hline &
&(ecqm)&(cqm)&(ecqm)&(cqm)&\\
\hline $\Sigma^{+}_{c}\rightarrow
\Lambda^{+}_{c}$&$-\frac{1}{\sqrt{3}}(\mu_{u}-\mu_{d})$&60.55&85.59&97.98&104.03&98.70\cite{JDey1994}\\
&&&&&&60.70\cite{Ivanov1998} \\
&&&&&&87.00\cite{Stawfiq2001} \\
&&&&&&89.00\cite{Fayyazuddin1997} \\
\hline $\Sigma^{*++}_{c}\rightarrow
\Sigma^{++}_{c}$&$-\frac{2\sqrt{2}}{3}(\mu_{u}-\mu_{c})$&1.15&1.71&1.98&2.22&1.70\cite{JDey1994}\\
&&&&&&3.04\cite{Stawfiq2001} \\
\hline $\Sigma^{*+}_{c}\rightarrow
\Sigma^{+}_{c}$&$\frac{\sqrt{2}}{3}(\mu_{u}+\mu_{d}-2\mu_{c})$&$0.00006$&$0.000079$&0.0112&0.0125&0.01\cite{JDey1994}\\
&&&&&&0.14\cite{Ivanov1998} \\
&&&&&&0.19\cite{Stawfiq2001} \\
\hline $\Sigma^{*0}_{c}\rightarrow
\Sigma^{0}_{c}$&$\frac{2\sqrt{2}}{3}(\mu_{d}-\mu_{c})$&1.12&1.66&1.44&1.60&1.20\cite{JDey1994}\\
\hline $\Sigma^{*+}_{c}\rightarrow
\Lambda^{+}_{c}$&$\frac{\sqrt{2}}{\sqrt3}(\mu_{u}-\mu_{d})$&154.48&229.85&244.39&273.48&250.00\cite{JDey1994}\\
&&&&&&151.00\cite{Ivanov1998} \\
&&&&&&230.00\cite{Fayyazuddin1997} \\
\hline $\Omega^{*0}_{c}\rightarrow
\Omega^{0}_{c}$&$\frac{\sqrt{2}}{3}(\mu_{s}-\mu_{c})$&2.02&3.13&0.82&0.79&0.36\cite{JDey1994}\\
\hline $\Xi^{*+}_{c}\rightarrow
\Xi^{+}_{c}$&$\frac{\sqrt{2}}{3}(\mu_{u}-\mu_{s})$&63.32&96.34&99.94&110.77&124.00\cite{JDey1994}\\
&&&&&&75.60\cite{Fayyazuddin1997} \\
\hline $\Xi^{*0}_{c}\rightarrow
\Xi^{0}_{c}$&$\frac{\sqrt{2}}{\sqrt{3}}(\mu_{u}-\mu_{s})$&0.30&0.46&1.15&1.27&0.80\cite{JDey1994}\\
&&&&&&0.90\cite{Fayyazuddin1997} \\
\hline
\end{tabular}
\end{center}
\end{table*}
We start with the color singlet Hamiltonian of the system  as
\begin{equation}\label{eq:2.01}
H=-\sum\limits_{i=1}^3\frac{\nabla_{i}^2}{2m_{i}}+\sum
\limits_{i<j} V_{ij}
\end{equation}
Where, the interquark potential\begin{center}$V_{ij}$=$ -\frac{2
\alpha_{s}}{3}\frac{1}{x_{ij}} +\beta \ x_{ij}
+V_{spin}(ij)$;\end{center} Here, $\alpha_{s}$ is the running
strong coupling constant, $\beta$ is the potential strength of the
baryonic system and $V_{spin}$ is the spin dependent part of the
two body system. For the present study, we considered
$\alpha_s(\mu_{0}=1 GeV)\approx 0.7$. All other parameters of the
model including the quark masses are obtained from the study of
mass spectra of the singly charmed baryons \cite{Ajay2007}.

\section{Radiative decay of single charmed baryons}

The electromagnetic radiative decay width can be expressed in
terms of the radiative transition magnetic moment(in $\mu_{N}$)
and photon energy (k) as \cite{JDey1994}
\begin{equation}
\Gamma_{\gamma}=\frac{k^3}{4\pi}\frac{2}{2J+1}\frac{e^2}{m_{p}^2}
\mu^2
\end{equation}
here, $m_{p}$  is the proton mass, $\mu$ is the radiative
transition magnetic moments (in nuclear magnetons), which are
expressed in terms of the magnetic moments of the constituting
quarks ($\mu_{q}$) of the initial state of the Baryon
\cite{JDey1994}. The magnetic moment of the constituting quarks
are obtained as \cite{Bhavin2008}
\begin{equation}
\mu_q=\left<\phi_{sf}\mid\frac{e_{i}}{2m_{q}^{eff}}
\overrightarrow{\sigma}_{i}\mid\phi_{sf}\right>
\end{equation}

where, $e_{i}$ and $\sigma_{i}$ represents the charge and the spin
of the quarks forming the baryonic state. We have employed the
spin flavour wavefunction ($\left|\phi_{sf}\right>$) of the
symmetric and antisymmetric states of the baryons as used in
\cite{Bhavin2008}. Here, $m_{q}^{eff}$ corresponds to  mass of the
bound quark inside the baryons taking in to account of its binding
interactions with other two quarks described by the Hamiltonian
given in Eqn \ref{eq:2.01}. The effective mass for each of the
constituting quark $m_{{q_{i}}}^{eff}$ can be defined as
\cite{Bhavin2008}
\begin{equation}
m_{q_{i}}^{eff}=m_i\left(
1+\frac{\left<H\right>}{\sum\limits_{i}m_{i}}\right) \\
\end{equation}
where, $\left<H\right>=E+\left<V_{spin}\right>$ such that the
corresponding mass of the baryon with spin angular momentum, J is
given by
\begin{equation}
M_B^{J}=\sum\limits_{i}m_{i}+\left<H\right>=\sum\limits_{i}m^{eff}_{i}\\
\end{equation}
Here, $m_{i}$'s are the model quark mass parameters.\\
Accordingly, the effective mass of the $u$ and $d$ quarks are
obtained from the baryonic states of udc. Using the mass spectra
and the magnetic moments of the charmed baryons, we compute the
transition decay widths with and without considering the effective
constituent masses of the quarks.  We have also computed the
transition widths using the predicted magnetic moments and masses
based on a hyper central model (HCM) \cite{Bhavin2008} for
comparision. \\
\section{Conclusion and discussion}
We have employed a simple nonrelativistic variational approach
with coulomb plus linear potential to compute the radiative decay
of the single heavy flavour baryons in terms of radiative
transition magnetic moments and photon energy. The model
parameters are obtained to get the ground state masses of the cqq
systems \cite{Ajay2007}.  The computed radiative transition widths
are listed in Table \ref{tab:01} and compared with other
theoretical models. Our results tabulated below are found to be in
accordance with other model predictions. It can also be seen that
our predictions with the effective mass of the quarks within the
baryons are in better agreement with the predictions of
ref.\cite{Ivanov1998}, while these without the effective mass
correction are close to the predictions of
ref.\cite{Fayyazuddin1997} in general. How ever the hypercentral
model predictions with the effective quark mass are in agreement
with that of ref.\cite{JDey1994}. The future experimental results
on these transition widths can only resolve these variations among
different model predictions.
\section{ Acknowledgements}
We acknowledge the financial support from University Grant
Commission, Government of India, under a Major Research Project
\textbf{F. 32-31/2006(SR)}.

\end{document}